\documentclass{aa}  

\usepackage{graphicx}
\usepackage{txfonts}
\usepackage{hyperref}

\usepackage{natbib}
\begin{document}

   \title{First optical validation of a Schwarzschild Couder telescope: the ASTRI SST-2M Cherenkov telescope}
   \titlerunning{Optical validation of the ASTRI SST-2M Cherenkov telescope}


   \authorrunning{E. Giro,  R. Canestrari \and G. Sironi et al.}
   \author{E. Giro\inst{1}$^,$\inst{2},
          R. Canestrari\inst{2},                        
         G. Sironi\inst{2},
         E. Antolini \inst{3}, 
         P. Conconi\inst{2}, 
         C.E. Fermino\inst{4},
         C. Gargano\inst{5},  
         G. Rodeghiero\inst{1}$^,$\inst{6}, 
         F. Russo\inst{7},
         S. Scuderi\inst{8},
         G. Tosti\inst{3},
         V. Vassiliev\inst{9},
         \and
         G. Pareschi\inst{2}
          }

   \institute{INAF Osservatorio Astronomico di Padova, vicolo dell' Osservatorio 5, 35122, Padova, ITALY\\
              \email{enrico.giro@oapd.inaf.it}
              \and
             INAF Osservatorio Astronomico di Brera, via Emilio Bianchi 46, 23807, Merate (LC), ITALY 
             \and            
             Dipartimento di Fisica e Geologia, Universit\'a di Perugia, via Pascoli, 06123, Perugia, ITALY
             \and
             Instituto de Astronomia, Geofisica e Ci{\^e}ncias Atmosfericas, Universidade de S{\~a}o Paulo, Rua do Mat{\~a}o 1226, S{\~a}o Paulo, BRASIL            
             \and
             INAF IASF Palermo, via Ugo La Malfa 153, 90146, Palermo, ITALY             
             \and
             Max Planck Institute for Astronomy, K{\"o}nigstuhl 17, 69117, Heidelberg, GERMANY    
             \and         
             INAF Osservatorio Astrofisico di Torino, via Osservatorio 20, 10025, Pino Torinese (TO), ITALY
             \and
             INAF Osservatorio Astrofisico di Catania, via S. Sofia 78, 95125, Catania, ITALY
         \and
             Department of Physics $\&$ Astronomy, University of California, 430 Portola Plaza, Los Angeles, CA 90095, USA             
             }

   \date{Received July 19, 2017; accepted September 20, 2017}

 
  \abstract
   {The Chernekov Telescope Array (CTA) represents the most advanced facility designed for Cherenkov Astronomy. ASTRI SST-2M has been developed as a demonstrator for the Small Size Telescope in the context of the upcoming CTA. Its main innovation consists in the optical layout which implements the Schwarzschild-Couder configuration and is fully validated for the first time. The ASTRI SST-2M optical system represents the first qualified example of a  two-mirror telescope for Cherenkov Astronomy. 
   This configuration permits us to (i) maintain  high optical quality across a large field of view,   (ii) demagnify the plate scale, and (iii) exploit new technological solutions for focal plane sensors.}
   {The goal of the paper is to present the optical qualification of the ASTRI SST-2M telescope. The qualification has been obtained measuring the point spread function (PSF) sizes generated in the focal plane at various distances from the optical axis. These values have been compared with the performances expected by design.}
   {After an introduction on  Gamma-ray Astronomy from the ground, the optical design of ASTRI SST-2M and how it has been implemented  is discussed. Moreover, the description of the set-up used to qualify the telescope over the full field of view is shown.}
   {We report the results of the first--light optical qualification. The required specification of a flat PSF of $\sim 10$ arcmin in a large field of view ($\sim 10^{\circ}$) has been demonstrated. These results validate the design specifications, opening a new scenario for Cherenkov Gamma-ray Astronomy and, in particular, for the detection of high-energy (5 - 300 TeV) gamma rays and wide-field observations with CTA.}
   {}


   \maketitle
%

\section{Introduction}

The German astrophysicist Karl Schwarzschild proposed a design for a two-mirror telescope that would eliminate much of the optical aberration across the field of view (FoV) \citep{Schwarzschild}. In his paper he proposed a method to design an optical system free of both spherical and coma aberrations by means of two aspherical mirrors described by radial polynomials. In recognition of Schwarzschild's contribution to  developments in optics,  the analytical solutions for two-mirror telescopes are known as `Schwarzschild aplanats'. In 1926 Couder enhanced Schwarzschild's solution by adding a curved focal plane to reduce  astigmatism \citep{Couder}, and this  configuration took the name of Schwarzschild--Couder (hereafter SC). More recently, Lynden-Bell \citep{Lynden-Bell}, starting from this pioneering work, proposed a solution for  exact optics in aplanatic telescopes.\\
In order to limit the coma aberration effects in X-ray microscopy, Wolter was the first who applied  Schwarzschild's approach to obtaining aplanatic grazing incidence optics for X-rays \citep{Woltera} and \citep{Wolterb}.  
Wolter's approach was later  adopted for the design of X--ray astronomical telescopes based on grazing incidence optics, and a number of X-ray telescopes were realized (\citealt{Aschenbacha} and \citealt{Aschenbachb}). Following the approach initially suggested by Burrows, Burgh and Giacconi \citep{Burrows}, Conconi \citep{Conconi} showed that different Wolter--like designs could be obtained by expanding as a power series the primary and secondary mirrors' profiles in order to increase the angular resolution at large off-axis positions, at the expense of the on-axis performances. Also, the curvature of the focal plane could be optimized, similarly to  Couder's solution for normal incidence mirrors.\\
However, for non-grazing incidence telescopes, this idea laid dormant for almost a century, consigned to suspended animation in specialized journals and texts on astrophysical optics because they were considered too difficult and expensive to build. Summarizing, it was the lack of technological solutions for aspherical mirrors that limited the applications of the Schwarzschild--Couder  design.\\
A new period in exploiting the hidden possibilities of the SC design was started by Vassiliev and his team with the design proposed for the Cherenkov telescopes in \citet{Vassiliev}. 
Eventually, the ASTRI Small Size Telescope dual Mirror (SST-2M) was the first telescope demonstrating that the SC configuration matches the optical requirements of Imaging Air Cherenkov Telescopes (IACTs).

IACTs are astronomical instruments capable of imaging very short light flashes. This effect is typical for Cherenkov radiation generated when a very high-energy gamma ray strikes the atmosphere (\cite{Galbraith} and \cite{Hinton}).\\
Gamma-ray photons, ranging from few GeV up to hundreds of TeV, interacting with the Earth atmosphere in the so-called pair production process generate an \textit{e$^+$e$^-$} couple that travels faster than light in the medium thus polarizing the molecules. At the de-excitation a faint flash of beamed visible near-UV photons is emitted (typical observation wavelength band 300 -- 600 nm\footnote{The 300-600 wavelength band is due to practical reasons; only a little light will arrive on the Earth's surface below 290 nm, while above 600 nm there are strong telluric lines \citep{2015A&A...576A..77S}}, a phenomenon  known as the Cherenkov effect \citealt{Cherenkov}). The \textit{e$^+$e$^-$} via the $bremsstrahlung$ process produces a lower energy gamma-ray photon and a further positron/electron and so on until $E_{c} = 83~{\rm MeV}$. IACTs collect the UV photons generated by the Cherenkov effect by means of suitable pixel sensors, such as  photomultiplier tubes. The image analysis yields information on the primary photons, such as the arrival direction and  energy. This analysis has been proposed by Hillas \citep{Hillas} and allows  gamma photons to be disentangled from other cosmic rays.\\
The IACTs realized so far  have simple optical configurations with a single reflecting mirror. A typical layout frequently adopted is the Davies-Cotton  (DC) design \citep{davies-cotton}, originally proposed for solar concentrators. The DC configuration was adopted for the HEGRA \citep{hegra}, H.E.S.S. \citep{2003APh....20..111B}, and VERITAS \citep{2008ApJ...679.1427A} Cherenkov telescopes. A second possible design is the parabolic configuration, adopted  for  the MAGIC experiment \citep{2016APh....72...61A}. Both these configurations are simple and cost-effective, but with the drawback of a relatively small corrected FoV for fast optical design (small F/$\#$). Their typical angular resolution is  about 3--10 arcmin across a FoV of  $2^{\circ} - 4^{\circ}$, with a typical focal ratio ranging from 0.7 to 1.2. This is suitable for observing sources in the band up to a few TeV. For higher energies, a FoV greater than $7^{\circ}$ is preferable.  Moreover, Davies-Cotton telescopes with apertures larger than 15 meters introduce non-negligible time delays, spoiling the isochronicity of the detected events.

Alternative configurations have been proposed.  \citet{Razmik} proposed for the first time an elliptical design and \citet{Mirzoyan}  proposed a solution based on a modified Schmidt with a weak aspherical mirror surface. In this second configuration the asphericity of the primary mirror is achieved by tilting spherical segments in order to approximate the desired geometry. The result is free of spherical aberrations and coma because a corrective lens is placed at the entrance pupil. The proposed design achieved a 2 arcmin of angular resolution across a corrected FoV of $15^{\circ}$ ~and a focal ratio of 0.8. The major drawbacks are the  large lens corrector that needs to be manufactured and the large physical dimensions of the focal surface.\\
Another configuration inspired by the Schmidt design has been proposed for the MACHETE experiment, but in this case the telescope is non-steerable \citep{Cortina}.\\
The solution proposed by \citet{Vassiliev} seems the most promising both in terms of fulfilling requirements and cost-effective implementation. The SC design permits  spherical, coma and astigmatism aberrations to be corrected for simultaneously.  Moreover, the use of the secondary mirror decreases the equivalent focal length to achieve an enlargement of the plate scale. Strictly speaking, this implementation demagnifies the image.  On the other hand,  the dual mirror configuration introduces a strong vignetting of the primary mirror so that only about 40\% of the primary surface works for collecting light for each position on the focal plane. Moreover, a dual mirror configuration implies a double reflection with a consequent loss of photons.\\
This optical configuration became very attractive for Cherenkov telescope implementations with the new possibilities opened by Silicon PhotoMultipliers (SiPMs), today more reliable in comparison with the photomultiplier tubes.
Having typical lateral dimension of few millimetres, a SiPM is the perfect solution for realizing  a compact Cherenkov camera with a large FoV, when coupled with  demagnifying optics.
On the other hand, this kind of solution requires manufacturing of more accurate optics. In fact, the requirement of maintaining the point spread function (PSF) dimension within one Cherenkov camera pixel becomes stricter with the plate scale. For this reason, the asphericity of the reflecting surfaces, differently from the solution proposed by \citet{Mirzoyan}, is built into the mirrors themselves by means of a suitable manufacturing process. \\
A detailed discussion on the advantage of the SC solution in sampling the focal plane of IACTs can be found in \citet{Rodeghiero}.
In the following sections we present the optical layout adopted for the  ASTRI SST-2M telescope and the results obtained across in the entire FoV in comparison with the theoretical design.

\section{ASTRI SST-2M optical design}

%

\label{sec:des}
The optical system (OS) design of ASTRI SST-2M was developed in compliance with the requirements of the Small Size Telescopes of the CTA observatory \citep{Acharya} by the Italian National Institute for Astrophysics (INAF). It consists of three optical surfaces: the primary mirror (M1), the secondary mirror (M2), and the curved focal surface (FS) populated with the silicon photon multiplier (SiPM) sensors. The effective focal length of the OS is $F=2141$ mm. The distance between M1 and M2 is $3108.4$ mm $=F/q$, where $q=0.6888$ is the first Schwarzschild aplanat parameter. The distance between M2 and FS is $519.6$ mm $=F(1-\alpha)$, where $\alpha=0.7573$ is the second parameter, which together with q defines the Schwarzschild aplanatic solution. It has been shown in 
\cite{Vassiliev} that the optical systems in the vicinity of $(q=2/3, \alpha=2/3)$ are nearly optimal for applications in ground-based gamma-ray astronomy. The aperture of ASTRI SST-2M is $D=4306$ mm, which makes its OS very fast, with F/D = 0.50. The other OS parameters of importance for this publication are summarized in Table \ref{tab:opt_param}.\\ 

\begin{table}
        \centering
        \caption{Subset of ASTRI SST-2M OS parameters.}
        \label{tab:opt_param}
        \begin{tabular}{lc} 
                \hline
                Telescope OS Parameters &  Value\\
                \hline
                Effective focal length &  2141 mm\\
                Working f/$\#$ & 0.5\\
                FoV & 9.6$^{\circ}$ \\
                Obscuration ratio & 40\% \\
                Plate scale & 37.53 mm/$^{\circ}$\\             
                \hline
        \end{tabular}
\end{table}

The camera of ASTRI SST-2M, which is being developed based on the CTA requirements by the ASTRI collaboration \citep{2014SPIE.9147E..0DC}, will have $7$ mm $\times$ $7$ mm SiPM pixels and cover 9.6$^{\circ}$ field of view. Given the plate scale of 37.53 mm/$^{\circ}$ of the OS, each pixel is equivalent to 0.19$^{\circ}$ $\times$ 0.19$^{\circ}$~on sky. The OS of ASTRI SST-2M was designed to provide correction of aberrations within the full camera field of view. The sag $z(r)$ of each of the three optical surfaces of the OS was conventionally parameterized by the sphere with the polynomial representing aspheric corrections,

\begin{center}

\begin{equation}
 z(r) = \frac{cr^{2}}{1+\sqrt{1-c^{2}r^{2}}} + \sum_{i=2}^N \alpha_{i} \left( \frac{r}{r_{\max }} \right) ^{2i}
,\end{equation}

\vspace{0.5cm}

\end{center}
where $r$ is the radial coordinate and $r_{\max }$ is its maximum value, $c$ is the curvature of the sphere, and $\alpha_{i}$ are the coefficients of the higher order aspheric correction terms. \\
The diameter of the circle corresponding to the encircled energy of 80$\%$, D80 parameter, was used as a figure of merit for the optimization of $\alpha_{i}$.  Similarly the encircled energy (EE) is the percentage of focused energy contained within a circular area with the diameter corresponding to the SiPM pixel size.
It was required in the optimization process that the EE should be larger than 80\% across the full FoV. \\
The effects of mirror fabrication and misalignment errors on the PSF, D80, and EE were not taken into account during the optimization of the $\alpha_{i}$ coefficients, assuming ideal OS implementation. 


This approach was previously adopted to optimize the PSF across the FoV of X-ray grazing incidence  telescopes by \citet{Conconi}. It follows a method that is  similar to the Ritchey-Chretien design since on-axis aberrations are admitted in order to cancel off-axis aberrations and obtaining in that way a PSF response flat across the entire  FoV. This can be considered a customization of the SC design for Cherenkov telescopes purposes \citep{Sironi3}.
 
The optimization has been performed using ZEMAX\footnote{ ZEMAX LLC, ZEMAX software, www.zemax.com} software. Table \ref{tab:coeff} shows the optimized aspheric $\alpha_{i}$ coefficients, which describe the figures of M1, M2, and FS.\\

\begin{table}

        \centering

        \caption{Parameters of the ASTRI SST-2M  optical design. For each surface we report the radial dimension, radius of curvature and coefficients of asphericity.}        

        \label{tab:coeff}

        \begin{tabular}{cccc} 

                \hline

                 &  M1 & M2  & Detector\\
                \hline

                $r_{max} $ &  2153 mm& 900 mm& 180 mm \\
                $1/c $ &  -8223 mm & 2180 mm & 1060 mm \\

                \hline

                $\alpha_{1}$ &  $0.0$ &$ 0.0$& $0.0$ \\

                $\alpha_{2}$ &  $2.06503e^{+01}$ & $1.06338e^{+01}$& $0.0$ \\

                $\alpha_{3}$ &  $-5.63245e^{+00}$ & $-1.53897e^{+01}$&$ 0.0$ \\

                $\alpha_{4}$ &  $3.13020e^{+00}$ & $3.71653e^{+00}$ &$0.0$\\

                $\alpha_{5}$ &  $8.33769e^{+00}$ & $1.16757e^{+00}$ & 0.0\\

                $\alpha_{6}$ &  $5.23834e^{+00}$ & $-2.91922e^{-01}$& 0.0 \\

                $\alpha_{7}$ &  $-1.37544e^{+00}$ & $-1.54081e^{-01}$ & 0.0\\

                $\alpha_{8}$ &  $-9.36096e^{+00}$ & $-5.68038e^{-02}$ & 0.0\\

                $\alpha_{9}$ &  $-6.10074e^{+00}$ & $4.76042e^{-02}$ & 0.0\\

                $\alpha_{10}$ & $1.25307e^{+01}$ & $-4.51272e^{-03}$ & 0.0\\            

                \hline

        \end{tabular}

\end{table}

\begin{figure}  
   \centering
        \includegraphics[scale=1., trim={0cm 0cm 0cm 0cm}, clip]{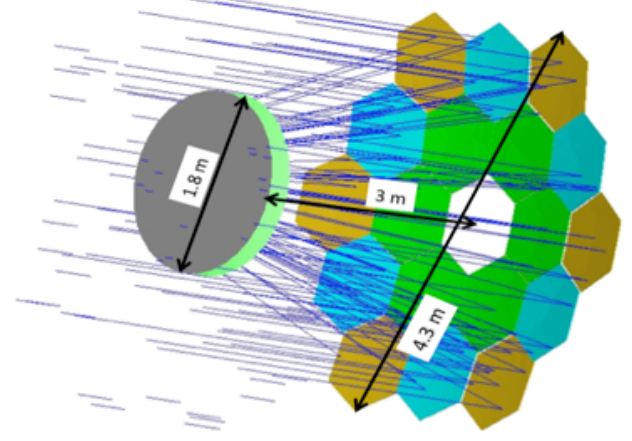}
    \caption{Optical layout for the ASTRI SST-2M telescope. The panels belonging to the three different rings are displayed in different colours.}
    \label{fig:opt_layout}
\end{figure}

It should be noted that the optimized radial profiles, expressed by the $\alpha_{i}$ coefficients, for both M1 and M2 are strongly aspheric. The practical realization of the M1 and M2 mirrors therefore represented a challenge, which was resolved with the ASTRI SST-2M OS implementation shown in Figure \ref{fig:opt_layout}. The M1 optical surface was segmented in a mosaic of 18 hexagonal panels of 849 mm which were distributed  among three concentric rings. Each ring was characterized by its radial distance from the optical axis and defines a different type of mirror segment with a free-form figure and asphericity $dz$ (referring to the best fitting sphere) of up to 1 mm. The cold glass slumping technology selected for the manufacturing of the M1 segments is described in \cite{Pareschi} and \cite{2013OptEn..52e1204C}. For quality control and characterization of the aspheric M1 segments, we developed a metrology testing method based on deflectometry, which is described in \cite{Sironi}. Unlike the segmented M1, the M2 mirror is monolithic. It was manufactured via the hot glass slumping technique \citep{Ghigo}. The qualification of M2 was performed by means of optical tests as explained in \citep{Rodeghiero}. The quality of the M1 and M2 manufacturing was verified with an ad hoc ray tracing software taking into account the measured surface of both mirrors and their alignment \citep{Sironi2}.
\begin{figure}
   \centering
   \includegraphics[scale=0.06, trim={0cm 1cm 0cm 5cm}, clip]{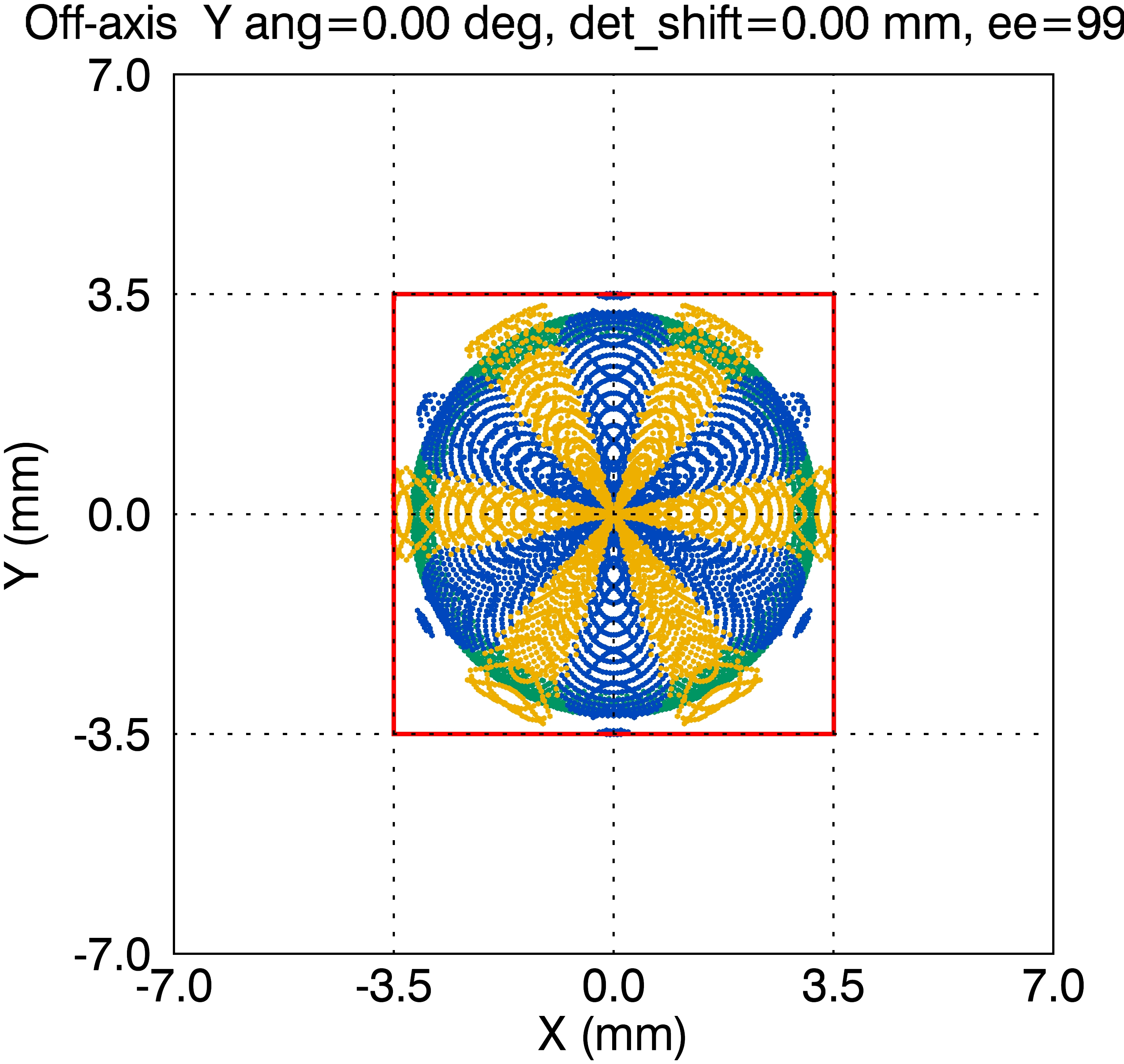}
    \caption{On-axis PSF of the ASTRI SST-2M telescope. The colour-coding of the contributions of the different segment types is  as in Figure \ref{fig:opt_layout}. The red square represents the dimension of the ASTRI camera pixel.}
    \label{fig:PSF_theo}
\end{figure}

\section{ASTRI SST-2M on-sky optical qualification}
The ASTRI SST-2M telescope is installed at the INAF `M.C. Fracastoro' observing station located in Serra La Nave (Mt. Etna), Italy, at an elevation of 1740 m above sea level. Its optical qualification was carried out using Polaris as a target and included two phases: 
\begin{itemize}
        \item[-] Alignment of the OS components;
        \item[-] PSF characterization over the entire FoV.
\end{itemize}

\subsection{Alignment of the OS components}
\begin{figure*}
        \includegraphics[scale=0.6, trim={0cm 0.cm 0cm 2cm}, clip]{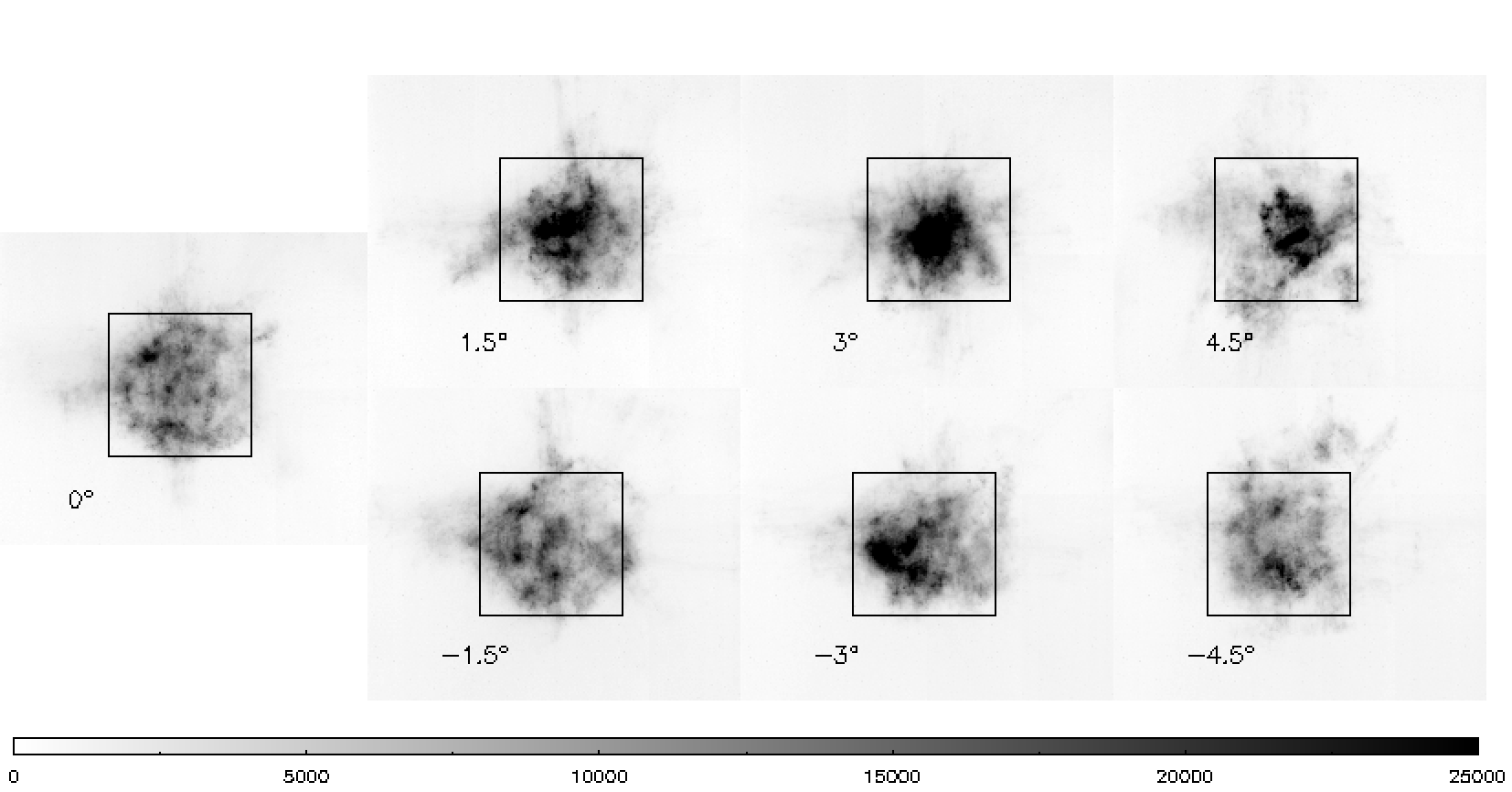}
        \vspace{0.2cm}
        \caption{PSF of the ASTRI SST-2M telescope across the focal plane. Alignment and optical performance have been optimized at $3^\circ$. The Cherenkov camera pixel size is overplotted for each PSF. }
    \label{fig:psf_fov}
\end{figure*}

The active mirror control system (AMC)~\citep{Gardiol} of ASTRI SST-2M consists of a tip-tilt correction alignment mechanism installed on each individual M1 segment and a tip-tilt combined with a translation alignment mechanism installed on the monolithic M2. Each M1 segment has a three-point mounting  with one point fixed on a ball joint and the other two attached to the actuators allowing tip-tilt adjustment in the range of $1.5^{\circ}$ ~with a resolution of a few arcseconds. Three actuators control the position of M2 permitting $15$mm motion along the optical axis with a resolution of $0.1$ mm and a maximum range of tip-tilt correction of $0.25^{\circ}$. 
To align the OS, a large  CCD (37 mm $\times$ 37 mm, corresponding to $1^{\circ}\times$ $1^{\circ}$ ~in the sky)\footnote{CCD KAF-09000 is mounted on a Pro Line Fingerlake camera} was placed at the nominal position of the telescope focal plane pointing at the Polaris star. The image was then analysed to deduce alignment corrections for M2 and each individual M1 segment.    
As described in section~\ref{sec:des}, the on-axis image of a point source for the ideal realization of ASTRI SST-2M OS has an extended image whose structure depends on the alignment of the OS components. Figure~\ref{fig:PSF_theo} shows the on-axis simulated structure of such an ideal image in which different colours  denote the contributions from the  rings illustrated in Figure~\ref{fig:opt_layout}. The alignment procedure consists of moving all the OS elements to reproduce the structure of the ideal PSF. As a first step in the alignment process the distance between M2 and the focal surface has been adjusted to match the dimension of the images of M1 segments of ideal on-axis PSF. Once this distance is fixed, the tip-tilt corrections were applied to each individual M1 segment to reproduce its contribution to the ideal PSF. Non-idealities of the fabrication of figures of individual M1 segments and M2 as well as the M1-to-M2 distance misalignment residual were not taken into account at this stage. A detailed description of all alignment procedures undertaken can be found in Sironi et al. (in prep.).\\
As a second step, a smaller size CCD ($27$ mm $ \times18$ mm i.e. $0.5^{\circ}\times  0.7^{\circ}$ in the sky)\footnote{CCD KAI-16050-A is assembled in a SVS-VISTEK GigE camera} has been mounted on a custom jig designed to allow the positioning of the camera at seven field angles on the focal surface: $0^{\circ}, \pm1.5^{\circ}, \pm3^{\circ}, \pm4.5^{\circ}$. A fine tuning of the tip-tilt corrections of the  M1 segments was performed at the off-axis position where the PSF dimension is minimized by design. As shown in Fig.  \ref{fig:psf_fov} this condition corresponds to $3^{\circ}$ off-axis.
\label{sec:align}

 \begin{figure}
     \includegraphics[scale=0.45, trim={1 0.5cm 1 0}, clip]{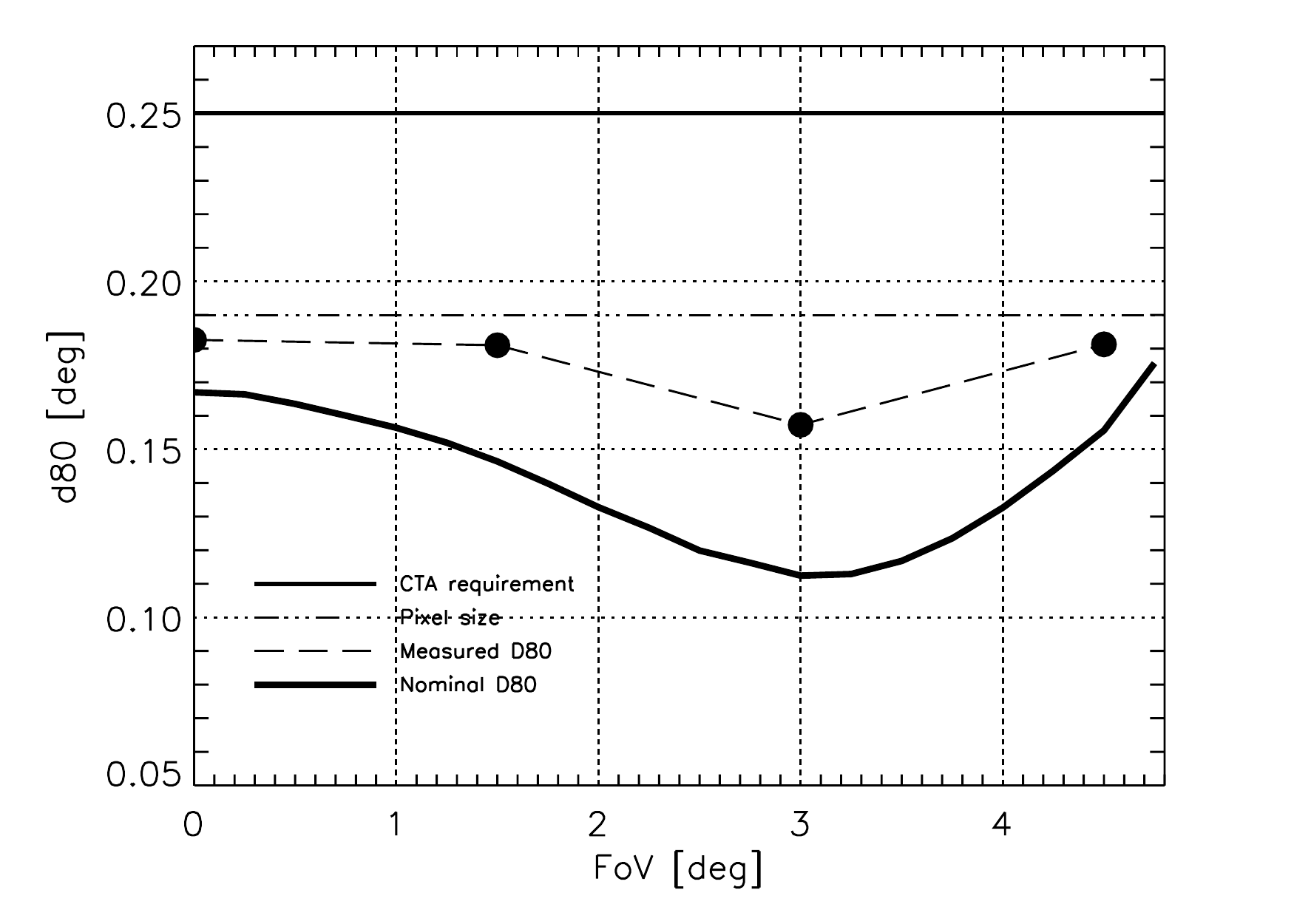}
    \caption{Comparison of D80 as a function of field angle (off-axis) between its ideal (solid curve) and measured at ASTRI SST-2M (dashed curve) values. The SiPM pixel size (dashed horizontal line) and CTA design requirement (solid horizontal line) are also shown for reference. The pixel size of the CCD camera used for these measurements is about $2$ arcsec on the focal plane.}
    \label{fig:d80}
\end{figure}

 \begin{figure}
   \includegraphics[scale=0.062, trim={15cm -3cm 20cm 1cm},clip]{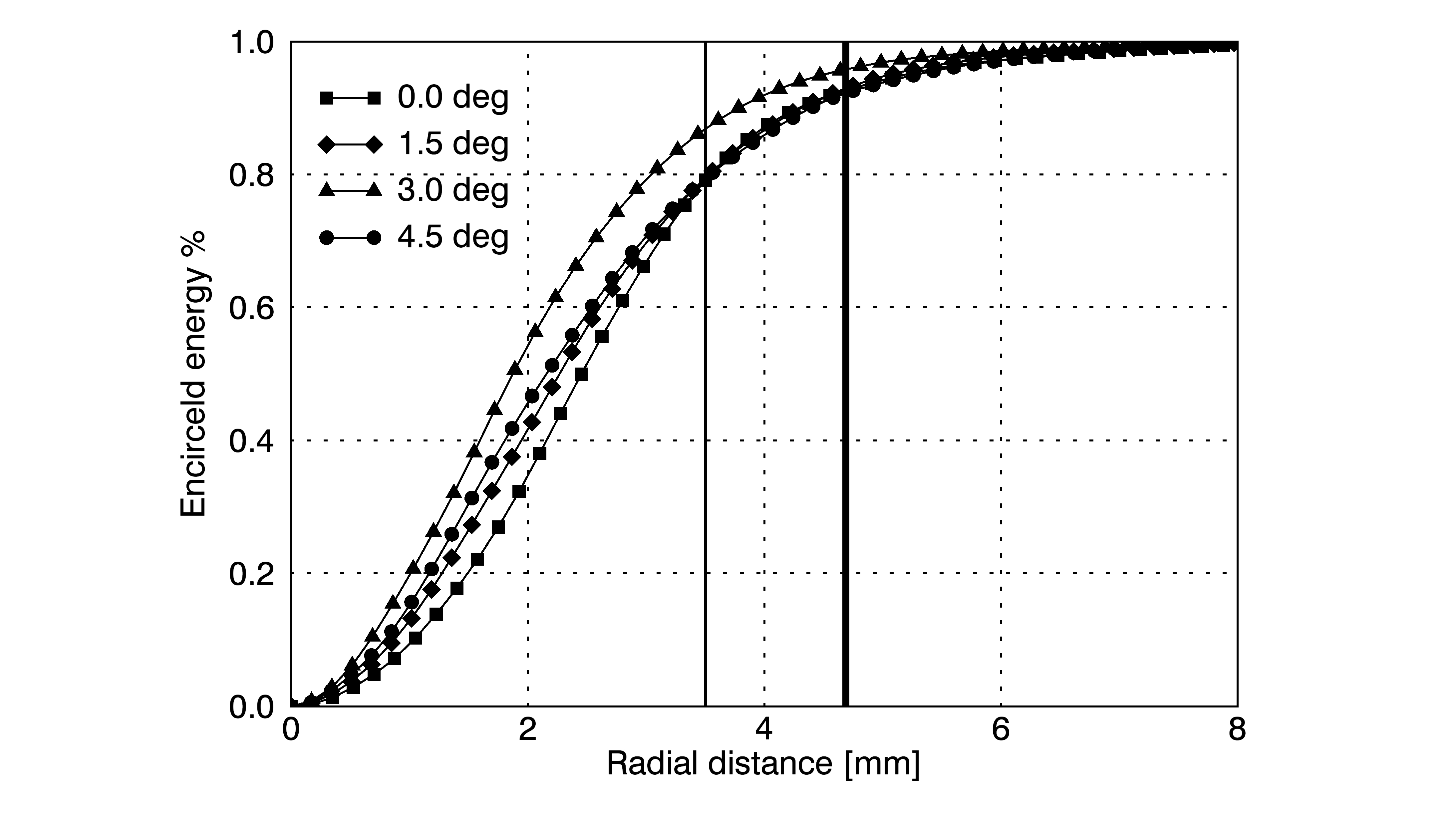}
    \vspace{-0.5cm}
    \caption{EE parameter measured as a function of radial distance for different angular positions on the focal plane. The vertical lines represent the SiPM pixel size (thin) and the CTA design requirement (thick).}
    \label{fig:d80_plot}
\end{figure}

\subsection{PSF characterization over the entire FoV}
To characterize PSF over the entire FoV we maintained the same configuration described for the fine-tuning. A typical PSF image covers an area with a diameter of about 600 CCD pixels reliably resolving details of the substructures in it.  For each of the seven field angles, four frames were acquired with and exposure of  2.5 seconds  on October 16, 2016. These images are shown in Fig. \ref{fig:psf_fov}. A square, representing the dimension of the SiPM pixel, is overplotted on each PSF for reference.
The energy flux in the small substructures of the PSF image visible outside the SiPM pixel area exceeds the total detected energy flux by a small amount.
Point spread functions look different across symmetric positions in the field, in particular those  corresponding to $\pm 4.5^{\circ}$. It should be noted  that the fine tuning alignment has been performed for the position corresponding to $3^{\circ}$ off-axis as reported in section \ref{sec:align}. This  implies that PSFs closer to this focal plane position are sharper than the others. The difference between images taken at the same angles is no more than 6\%.

To calculate the D80 parameter, a background image was acquired on the nearby sky patch without bright stars in the CCD field and it was directly subtracted from the corresponding PSF image. The D80 values for each off-axis position were calculated using the total photon flux contained in the subtracted PSF image of the entire CCD field and then plotted in Figure~\ref{fig:d80}. 
The obtained values (filled circles on dashed curve) follow a pattern similar to the ideal PSF design (continuous curve) with an additional aberration due to contributions from manufacturing errors of M1 and M2 figures and residual misalignment of the OS. These contributions are consistent with the estimates foreseen during the ASTRI SST-2M design for uncorrelated components and are  about $0.1^{\circ}$.\\
In summary, the measured D80 values are contained within the SiPM pixel size (dash-dotted horizontal line) across the entire FoV and they are well below the CTA requirement (horizontal solid line). Figure \ref{fig:d80_plot} shows the EE values derived from the subtracted PSF images for each off-axis position. \

\section{Conclusions}
With the advent of the Cherenkov Telescope Array a huge effort in the design of the next generation of IACTs has been carried out by the scientific community.\\
In this paper, after a brief description of Cherenkov observations and their role in high-energy astrophysics, the most common optical solutions adopted for this kind of telescope have been introduced.
Moreover, we described the  optical designs used to optimize the PSF quality in wide-field configurations.
The ASTRI SST-2M telescope implements an optical layout based on the Schwarzschild-Couder solution. Even though the mathematical description of the SC layout was  presented about 100 years ago, this is the first time that such a telescope has been built and optically tested. The results of its characterization demonstrated that the optical quality of the implemented telescope across the FoV fulfils the requirements for Cherenkov observations at the higher energies (> 1 TeV), showing a good agreement with the expected performances.\\

\begin{acknowledgements}
 The authors acknowledge the support of the local staff of the Serra La Nave observing station, the INAF-Osservatorio Astrofisico di Catania, and  the whole ASTRI team.
They also acknowledge companies involved in the realization of the telescope, in particular the GEC consortium (composed of EIE and the Galbiati group), MediaLario Technologies, Flabeg, and Zaot.\\
This work is supported by the Italian Ministry of Education, University, and Research (MIUR) with funds specifically assigned to the Italian National Institute of Astrophysics (INAF) for the Cherenkov Telescope Array (CTA), and by the Italian Ministry of Economic Development (MISE) within the ``Astronomia Industriale'' programme. We acknowledge support from  the Brazilian Funding Agency FAPESP (Grant 2013/10559-5) and from the South African Department of Science and Technology through Funding Agreement 0227/2014 for the South African Gamma-Ray Astronomy Programme.
R. Canestrari and G. Sironi also acknowledge  the the support  from the Grant ``Cariplo/Regione Lombardia ID 2014-1980/ RST - BANDO congiunto Fondazione Cariplo-Regione Lombardia - ERC'' to the project ``Science and Technology at the frontiers of Gamma-Ray Astronomy with imaging atmospheric Cherenkov telescopes''.
\end{acknowledgements}

This paper has gone through internal review by the CTA Consortium.%


\end{document}